\documentclass[aps,prl,reprint,preprintnumbers,floatfix,superscriptaddress]{revtex4-1}
\usepackage{graphicx}
\usepackage{xspace}
\usepackage{SIunits}

\newcommand{\nue}{\ensuremath{\nu_{e}}\xspace}

\newcommand{\nuebar}{\ensuremath{\bar\nu_{e}}\xspace}

\newcommand{\numu}{\ensuremath{\nu_{\mu}}\xspace}

\newcommand{\numubar}{\ensuremath{\bar\nu_{\mu}}\xspace}

\usepackage{verbatim}

\usepackage{graphicx}
\usepackage{dcolumn}
\usepackage{bm}
\usepackage{amsmath}

\usepackage{hyperref}
\usepackage{url}
\usepackage{cleveref}
\Crefformat{figure}{#2Fig.~#1#3}
\Crefmultiformat{figure}{Figs.~#2#1#3}{ and~#2#1#3}{, #2#1#3}{ and~#2#1#3}

\usepackage[font=small,labelfont=bf,labelsep=space]{caption}
\usepackage{lineno}

\begin{document}
\preprint{FERMILAB-PUB-22-446-ND}
\title{
    Measurement of the $\nu_e-$Nucleus Charged-Current \\ Double-Differential Cross Section at $\left< E_{\nu} \right> = $ \unit{2.4}{\giga \electronvolt} using NOvA
}

\newcommand{\ANL}{Argonne National Laboratory, Argonne, Illinois 60439, 
USA}
\newcommand{\ICS}{Institute of Computer Science, The Czech 
Academy of Sciences, 
182 07 Prague, Czech Republic}
\newcommand{\IOP}{Institute of Physics, The Czech 
Academy of Sciences, 
182 21 Prague, Czech Republic}
\newcommand{\Atlantico}{Universidad del Atlantico,
Carrera 30 No. 8-49, Puerto Colombia, Atlantico, Colombia}
\newcommand{\BHU}{Department of Physics, Institute of Science, Banaras 
Hindu University, Varanasi, 221 005, India}
\newcommand{\UCLA}{Physics and Astronomy Department, UCLA, Box 951547, Los 
Angeles, California 90095-1547, USA}
\newcommand{\Caltech}{California Institute of 
Technology, Pasadena, California 91125, USA}
\newcommand{\Cochin}{Department of Physics, Cochin University
of Science and Technology, Kochi 682 022, India}
\newcommand{\Charles}
{Charles University, Faculty of Mathematics and Physics,
 Institute of Particle and Nuclear Physics, Prague, Czech Republic}
\newcommand{\Cincinnati}{Department of Physics, University of Cincinnati, 
Cincinnati, Ohio 45221, USA}
\newcommand{\CSU}{Department of Physics, Colorado 
State University, Fort Collins, CO 80523-1875, USA}
\newcommand{\CTU}{Czech Technical University in Prague,
Brehova 7, 115 19 Prague 1, Czech Republic}
\newcommand{\Dallas}{Physics Department, University of Texas at Dallas,
800 W. Campbell Rd. Richardson, Texas 75083-0688, USA}
\newcommand{\DallasU}{University of Dallas, 1845 E 
Northgate Drive, Irving, Texas 75062 USA}
\newcommand{\Delhi}{Department of Physics and Astrophysics, University of 
Delhi, Delhi 110007, India}
\newcommand{\JINR}{Joint Institute for Nuclear Research,  
Dubna, Moscow region 141980, Russia}
\newcommand{\Erciyes}{
Department of Physics, Erciyes University, Kayseri 38030, Turkey}
\newcommand{\FNAL}{Fermi National Accelerator Laboratory, Batavia, 
Illinois 60510, USA}
\newcommand{\UFG}{Instituto de F\'{i}sica, Universidade Federal de 
Goi\'{a}s, Goi\^{a}nia, Goi\'{a}s, 74690-900, Brazil}
\newcommand{\Guwahati}{Department of Physics, IIT Guwahati, Guwahati, 781 
039, India}
\newcommand{\Harvard}{Department of Physics, Harvard University, 
Cambridge, Massachusetts 02138, USA}
\newcommand{\Houston}{Department of Physics, 
University of Houston, Houston, Texas 77204, USA}
\newcommand{\IHyderabad}{Department of Physics, IIT Hyderabad, Hyderabad, 
502 205, India}
\newcommand{\Hyderabad}{School of Physics, University of Hyderabad, 
Hyderabad, 500 046, India}
\newcommand{\IIT}{Illinois Institute of Technology,
Chicago IL 60616, USA}
\newcommand{\Indiana}{Indiana University, Bloomington, Indiana 47405, 
USA}
\newcommand{\INR}{Institute for Nuclear Research of Russia, Academy of 
Sciences 7a, 60th October Anniversary prospect, Moscow 117312, Russia}
\newcommand{\Iowa}{Department of Physics and Astronomy, Iowa State 
University, Ames, Iowa 50011, USA}
\newcommand{\Irvine}{Department of Physics and Astronomy, 
University of California at Irvine, Irvine, California 92697, USA}
\newcommand{\Jammu}{Department of Physics and Electronics, University of 
Jammu, Jammu Tawi, 180 006, Jammu and Kashmir, India}
\newcommand{\Lebedev}{Nuclear Physics and Astrophysics Division, Lebedev 
Physical 
Institute, Leninsky Prospect 53, 119991 Moscow, Russia}
\newcommand{\Magdalena}{Universidad del Magdalena, Carrera 32 No 22-08 Santa Marta, Colombia}
\newcommand{\MSU}{Department of Physics and Astronomy, Michigan State 
University, East Lansing, Michigan 48824, USA}
\newcommand{\Crookston}{Math, Science and Technology Department, University 
of Minnesota Crookston, Crookston, Minnesota 56716, USA}
\newcommand{\Duluth}{Department of Physics and Astronomy, 
University of Minnesota Duluth, Duluth, Minnesota 55812, USA}
\newcommand{\Minnesota}{School of Physics and Astronomy, University of 
Minnesota Twin Cities, Minneapolis, Minnesota 55455, USA}
\newcommand{\Mississippi}{University of Mississippi, University, Mississippi 38677, USA}
\newcommand{\NISER}{National Institute of Science Education and Research,
Khurda, 752050, Odisha, India}
\newcommand{\Oxford}{Subdepartment of Particle Physics, 
University of Oxford, Oxford OX1 3RH, United Kingdom}
\newcommand{\Panjab}{Department of Physics, Panjab University, 
Chandigarh, 160 014, India}
\newcommand{\Pitt}{Department of Physics, 
University of Pittsburgh, Pittsburgh, Pennsylvania 15260, USA}
\newcommand{\QMU}{Particle Physics Research Centre, 
Department of Physics and Astronomy,
Queen Mary University of London,
London E1 4NS, United Kingdom}
\newcommand{\RAL}{Rutherford Appleton Laboratory, Science 
and 
Technology Facilities Council, Didcot, OX11 0QX, United Kingdom}
\newcommand{\SAlabama}{Department of Physics, University of 
South Alabama, Mobile, Alabama 36688, USA} 
\newcommand{\Carolina}{Department of Physics and Astronomy, University of 
South Carolina, Columbia, South Carolina 29208, USA}
\newcommand{\SDakota}{South Dakota School of Mines and Technology, Rapid 
City, South Dakota 57701, USA}
\newcommand{\SMU}{Department of Physics, Southern Methodist University, 
Dallas, Texas 75275, USA}
\newcommand{\Stanford}{Department of Physics, Stanford University, 
Stanford, California 94305, USA}
\newcommand{\Sussex}{Department of Physics and Astronomy, University of 
Sussex, Falmer, Brighton BN1 9QH, United Kingdom}
\newcommand{\Syracuse}{Department of Physics, Syracuse University,
Syracuse NY 13210, USA}
\newcommand{\Tennessee}{Department of Physics and Astronomy, 
University of Tennessee, Knoxville, Tennessee 37996, USA}
\newcommand{\Texas}{Department of Physics, University of Texas at Austin, 
Austin, Texas 78712, USA}
\newcommand{\Tufts}{Department of Physics and Astronomy, Tufts University, Medford, 
Massachusetts 02155, USA}
\newcommand{\UCL}{Physics and Astronomy Department, University College 
London, 
Gower Street, London WC1E 6BT, United Kingdom}
\newcommand{\Virginia}{Department of Physics, University of Virginia, 
Charlottesville, Virginia 22904, USA}
\newcommand{\WSU}{Department of Mathematics, Statistics, and Physics,
 Wichita State University, 
Wichita, Kansas 67206, USA}
\newcommand{\WandM}{Department of Physics, William \& Mary, 
Williamsburg, Virginia 23187, USA}
\newcommand{\Wisconsin}{Department of Physics, University of 
Wisconsin-Madison, Madison, Wisconsin 53706, USA}
\newcommand{\deceased}{Deceased.}
\affiliation{\ANL}
\affiliation{\Atlantico}
\affiliation{\BHU}
\affiliation{\Caltech}
\affiliation{\Charles}
\affiliation{\Cincinnati}
\affiliation{\Cochin}
\affiliation{\CSU}
\affiliation{\CTU}
\affiliation{\Delhi}
\affiliation{\Erciyes}
\affiliation{\FNAL}
\affiliation{\UFG}
\affiliation{\Guwahati}
\affiliation{\Houston}
\affiliation{\Hyderabad}
\affiliation{\IHyderabad}
\affiliation{\IIT}
\affiliation{\Indiana}
\affiliation{\ICS}
\affiliation{\INR}
\affiliation{\IOP}
\affiliation{\Iowa}
\affiliation{\Irvine}
\affiliation{\JINR}
\affiliation{\Lebedev}
\affiliation{\Magdalena}
\affiliation{\MSU}
\affiliation{\Duluth}
\affiliation{\Minnesota}
\affiliation{\Mississippi}
\affiliation{\NISER}
\affiliation{\Panjab}
\affiliation{\Pitt}
\affiliation{\QMU}
\affiliation{\SAlabama}
\affiliation{\Carolina}
\affiliation{\SMU}
\affiliation{\Sussex}
\affiliation{\Syracuse}
\affiliation{\Texas}
\affiliation{\Tufts}
\affiliation{\UCL}
\affiliation{\Virginia}
\affiliation{\WSU}
\affiliation{\WandM}
\affiliation{\Wisconsin}

\author{M.~A.~Acero}
\affiliation{\Atlantico}

\author{P.~Adamson}
\affiliation{\FNAL}



\author{L.~Aliaga}
\affiliation{\FNAL}






\author{N.~Anfimov}
\affiliation{\JINR}


\author{A.~Antoshkin}
\affiliation{\JINR}


\author{E.~Arrieta-Diaz}
\affiliation{\Magdalena}

\author{L.~Asquith}
\affiliation{\Sussex}


\author{A.~Aurisano}
\affiliation{\Cincinnati}


\author{A.~Back}
\affiliation{\Indiana}
\affiliation{\Iowa}

\author{C.~Backhouse}
\affiliation{\UCL}

% ...

\collaboration{The NOvA Collaboration}
\noaffiliation

\begin{abstract}
The inclusive electron neutrino charged-current cross section is measured in the NOvA near detector using $8.02\times10^{20}$ protons-on-target (POT) in
the NuMI beam. The sample of GeV electron neutrino interactions is the largest analyzed to date and is limited by $\simeq$ 17\% systematic rather than the $\simeq$ 7.4\% statistical uncertainties. The double-differential cross section in final-state electron energy and angle is presented for the first time, together with the single-differential dependence on $Q^{2}$ (squared four-momentum transfer) and energy, in the range \unit{1}{\giga \electronvolt} $ \leq E_{\nu} < $ \unit{6}{\giga \electronvolt}.  Detailed comparisons are made to the predictions of the GENIE, GiBUU, NEUT, and NuWro neutrino event generators. The data do not strongly favor a model over the others consistently across all three cross sections measured, though some models have especially good or poor agreement in the single differential cross section vs. $Q^{2}$.
\end{abstract}

\maketitle

Precision measurements of neutrino oscillation parameters~\cite{NOvA:2019cyt,PhysRevD.103.L011101,DUNE:2020lwj,Yokoyama:2017mnt}, such as  $\delta_{\rm{CP}}$ and the neutrino mass ordering use \nue (\nuebar ) appearance in predominantly \numu (\numubar) beams and depend on detailed understanding of neutrino interaction models. Despite the two-detector technique that largely mitigates the impact of cross-section uncertainties, model predictions are still required by oscillation measurements to estimate the energy spectrum and selection efficiencies of \nue interactions, requiring accurate models of interaction rate and outgoing-particle kinematics of \nue interactions.

Lepton universality suggests that \numu and \nue interactions are largely similar, differences in final-state lepton mass,  tree-level radiative corrections, and imprecisely known form factors of the nucleon reaction lead to differences in the predicted cross sections and their uncertainties \cite{ref:Day-McFarland}. There are few direct \nue cross-section measurements at the GeV energy scale as \nue are a small component ($\leq 1\%$) of the neutrino beams produced at accelerators. The dominant \numu component produces significant backgrounds. Previous measurements in this energy range have been performed by Gargamelle~\cite{ref:Gargamelle_nue} and  T2K~\cite{ref:T2K_nue,ref:T2K_Nue2020}, both of which reported the total integrated cross section as a function of energy.  MINERvA has reported single-differential cross sections for the quasi-elastic component of the $\nu_{e}$ - carbon cross section~\cite{ref:MINERvA_nue}. The measurement presented here uses the largest \unit{}{\giga \electronvolt} sample of electron neutrino-initiated interactions to date, with which the simultaneous differential dependence on electron energy, $E_e$, and angle, $\theta_e$, is extracted for the first time. The cross sections as a function of squared four-momentum transfer, $Q^{2}$, and as a function of neutrino energy, $E_{\nu}$, is also presented. 

Generators, such as~\cite{Andreopoulos:2009rq, Lalakulich:2011eh, Hayato:2009zz,Golan:2012rfa}, group the predictions of neutrino-nucleus interaction cross sections into three processes relevant in this energy range: quasielastic (QE) scattering, baryon resonance production (RES), and deep inelastic scattering (DIS). The modeling of these processes must be altered in the presence of a nuclear medium. A fourth process due to meson exchange current (MEC), occurs only inside the nuclei. These nuclear effects are not completely understood and the implementation of existing models in generators is currently incomplete \cite{PhysRevC.80.065501,PhysRevC.83.045501,Megias:2016fjk,Van_Cuyck_2017}. Measured differential dependences using electron neutrinos can be used to constrain the underlying interaction channels and elucidate nuclear effects. These measurements can also probe key aspects of neutrino-nucleus interaction models used to calculate oscillation-experiment acceptances.

\begin{figure}

\centering

\includegraphics[scale=0.432]{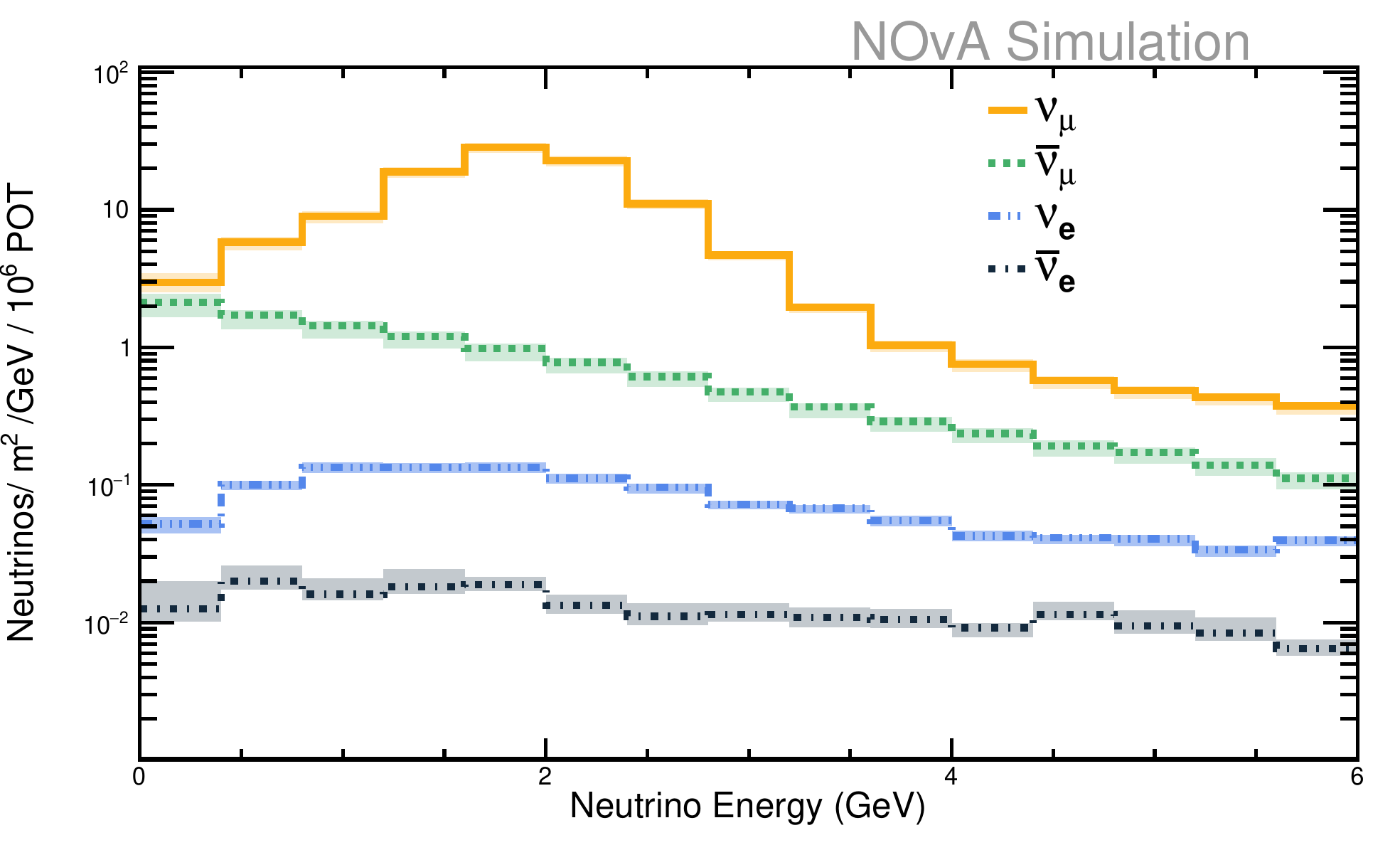}

\caption{ NUMI neutrino-mode beam component flux spectra at the NOvA near detector below \unit{6}{\giga \electronvolt}. The bands represent the total flux uncertainty on hadron production and beam optics for each component. }\label{fig:nue_flux}

\end{figure}

The NOvA experiment is designed to measure neutrino flavor oscillations~\cite{NOvA:2019cyt} using two detectors separated by \unit{809}{\kilo \meter}, placed \unit{14.6}{\milli \radian} off-axis from the central beam direction of Fermilab's NuMI beam~\cite{ref:Adamson:2015dkw}.  Magnetic focusing horns
in the beamline charge-select neutrino parents giving 96\%  pure \numu flux between 1 and \unit{6}{\giga \electronvolt}. The energy spectrum peak at \unit{1.8}{ \giga \electronvolt } at the near detector (ND) as shown in \Cref{fig:nue_flux}. The largest contamination is \numubar; the $<1\%$ $\nue + \nuebar$ arises mostly from muon decays below \unit{3}{ \giga \electronvolt} neutrino energy and kaon decays above.

The ND is a tracking calorimeter composed of liquid scintillator contained in PVC cells. The portion of the detector relevant to this measurement is \unit{3.9}{\meter}~$\times$~\unit{3.9}{\meter}~$\times$~\unit{12.8}{\meter} (long dimension along beam direction) in size and is segmented into cells that are \unit{6.6}{\centi \meter} $\times$ \unit{3.9}{\centi \meter} (0.15 radiation length $\times$ 0.45 Moli\`ere radius) in cross section and span the height and width of the detector in planes of alternating vertical and horizontal orientation. Each cell is filled with a blend of 95\% mineral oil and 5\% pseudocumene with trace concentrations of wavelength shifting fluors~\cite{ref:scintillator}. The resulting composition by mass is 67\% carbon, 16\% chlorine, 11\% hydrogen, 3\% titanium, 3\% oxygen with other trace elements. When a particle traverses the detector, wavelength shifting fiber in the cell collect and deliver scintillation light to avalanche photodiodes.  The result is digitized by custom front-end electronics.  All signals above a noise-vetoing threshold are sent to a data buffer. 
This Letter presents data corresponding to $8.02 \times10^{20}$ protons delivered to the NuMI production target (POT) between November 2014 and February 2017. 

This analysis relies on simulation to calculate the integrated flux, optimize event selection criteria, estimate selection efficiency, assess the effects of detector resolution and acceptance, and assess systematic uncertainties. The simulation proceeds in several steps: first neutrinos are generated from simulated mesons in the NuMI beam, then those neutrinos interact with nuclei in the detector, after which the final-state particles are transported through the detector.

The NuMI flux is predicted using Geant4 v9.03~\cite{AGOSTINELLI2003250} with the FTFP BERT hadronic model. The hadron production model is adjusted and constrained using external measurements by the PPFX package~\cite{Aliaga:2016oaz}.  Neutrino interactions are simulated using the GENIE v2.12.2~\cite{Andreopoulos:2009rq} event generator, hereinafter referred to as GENIE v2. The neutrino-interaction model output is adjusted to incorporate advances in theory and external data to better match a sample of \numu~charged-current(CC) interactions from NOvA ND data as detailed in Ref. ~\citep{ref:nova_xsec_tune}. In this dedicated tune, hereinafter referred to as the NOvA-tune, there are three substantial adjustments to the GENIE v2 prediction: (1) the QE axial mass ($M_{A}$ CCQE) is changed from 0.99 to \unit{1.04}{\giga \electronvolt \per c^{2}}~\cite{Meyer:2016oeg}; (2) soft non-resonant single pion production events from neutrinos are reduced by 57\%; (3) the empirical MEC model~\cite{Katori:2013eoa} is tuned to ND data using a two-dimensional fit in hadronic energy and momentum transfer. This procedure changes the shape of the MEC prediction and increases the cross section for that process by an average of 50\%. The NOvA-tune is applied to all simulated CC interactions. 

Geant4 v10.1.p3 is used to simulate energy deposited in the NOvA ND from the particles generated by neutrino interactions. Photon generation and propogation in the cells is modeled separately as follows: (1) the modeling and transfer of scintillation light uses NOvA measurements of scintillator response and fiber attenuation properties~\cite{ref:Aurisano_2015}; (2) the Birks suppression of the scintillation light is tuned using test-stand measurements and validated using a custom simulated response of the readout electronics~\cite{ref:Anfimov_2020}.

The signature of \nue CC interactions is an electron in the final state that produces an electromagnetic cascade within the detector. The major backgrounds are neutral current (NC) and \numu CC interactions with final-state $\pi^{0}$s that decay into two photons. These can mimic electron energy depositions if the showers overlap or one photon is low energy.

Candidate neutrino interactions (events) are reconstructed by forming collections of observed energy deposits (hits) from final-state charged particles correlated in space and time~\cite{10.5555/3001460.3001507}. Events are assigned a vertex by minimizing the angular spread of hits relative to candidate vertices~\cite{Baird:2015pgm}. Using a fuzzy K-means algorithm, hits within the event are clustered in regions of space emanating from the reconstructed interaction vertex into reconstructed particle trajectories (prongs)~\cite{doi:10.1080/01969727308546046}. Event primary vertices are required to be contained in a \unit{49}{\tonne} fiducial volume spanning \unit{2.8}{\meter} in both width and height and \unit{6.5}{\meter} in length. All prongs are required to stop at least \unit{30}{\centi \meter} from any edge of the detector to ensure containment. Events containing between 20 and 200 hits and at least one prong are selected for this analysis. Events outside this range are generally low visible energy NC events or high multiplicity events in which the electromagnetic cascade is difficult to isolate among the energy depositions from the other particles in the final state. 

\begin{figure}
\centering
\includegraphics[scale=0.432]{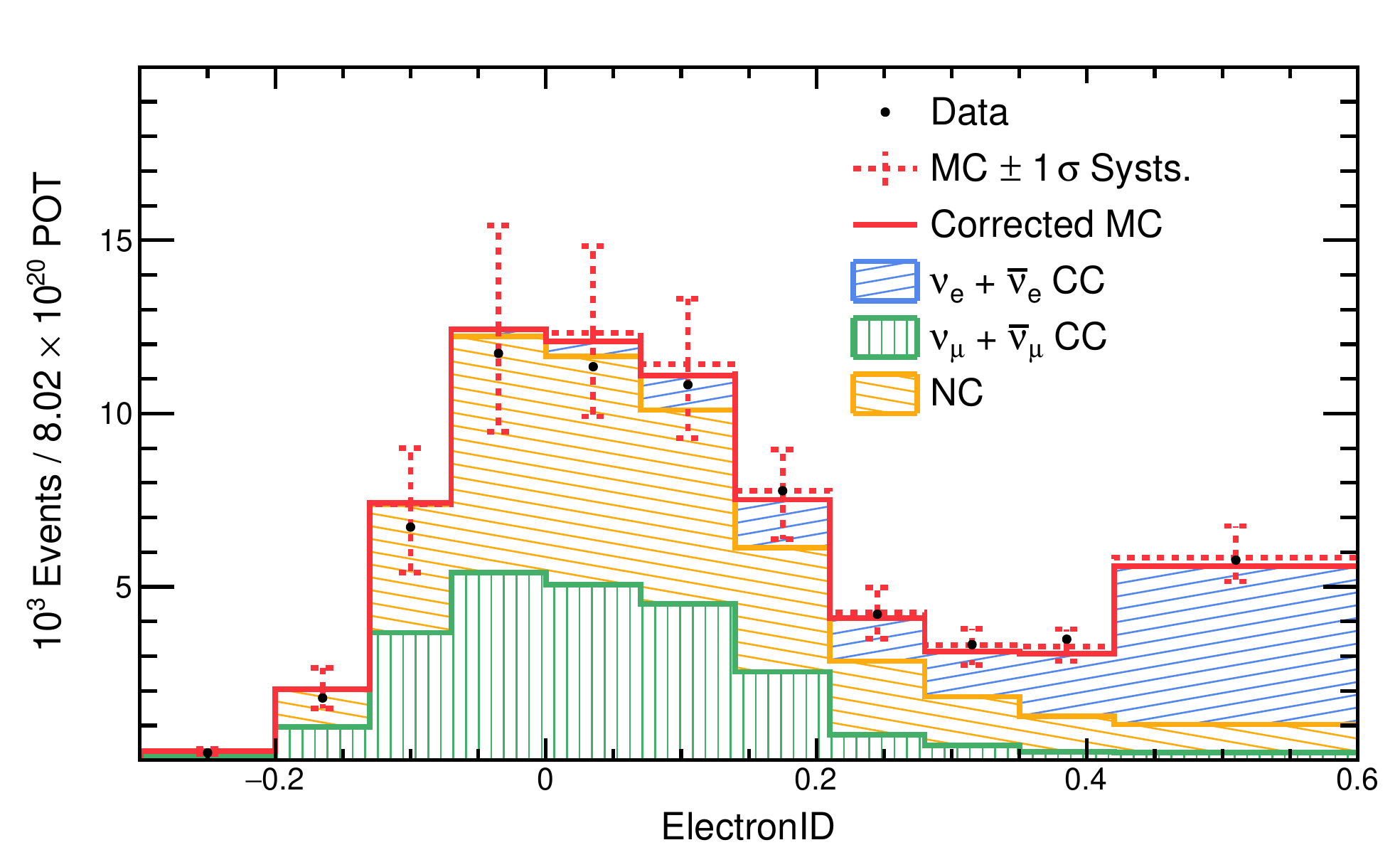}
\caption{Simulated ElectronID distributions of $\nu_{e} + \bar{\nu}_{e}$ CC (blue), $\nu_{\mu} + \bar{\nu}_{\mu}$ CC (orange), and NC (green), compared to data integrated over the reported electron kinematic phase space. The dashed (solid) line shows the total prediction from simulation (after extracted fit normalization corrections). Statistical uncertainties on data are too small to be seen. The vertical errors represents the $\pm 1 \sigma$ systematic range. All spectra are normalized to the data exposure.}
\label{fig:electronid}
\end{figure}

Prongs are identified as candidate electrons using a Boosted Decision Tree (BDT) algorithm~\cite{2007physics...3039H}. One input to the BDT is the output of a convolutional neural network (CNN) trained using simulated samples of single particles which labels prongs as muonic, electromagnetic, or hadronic in origin. Other inputs include the energy-weighted transverse width of the prong and the distance from the vertex to the prong start which aid in the separation of photon- from electron-induced showers. The performance of this BDT was validated against a sideband of data enriched with EM showers in $\nu_{\mu}$CC$\pi^{0}$ events. Candidate events are assigned an ElectronID score equal to the most electron-like BDT score of their constituent prongs. \Cref{fig:electronid} shows the ElectronID distribution in data and simulation after all selection criteria (and kinematic restrictions, defined below) are applied. 

The selected sample is binned according to the candidate electron shower's energy $E_{e}$, and its angle $\cos \theta_{e}$. Within each electron kinematic bin, the simulated ElectronID distribution is used to generate signal and background predictions, or templates. Each bin's templates are broken further into three components, whose normalizations are constrained in a simultaneous fit to the data: \nue+\nuebar, \numu+\numubar, and NC. Neutrinos and antineutrinos are not separated due to similarities in the shapes of their ElectronID templates. 

Electron-kinematic bins that satisfy two criteria are included in the fit: at least 100 predicted signal events and an estimated signal-to-background ratio $> 0.4$ in the ElectronID $> 0.2$ region. These requirements ensure adequate shape discrimination between the signal and background templates and limit the measured kinematic range to:

\begin{equation}
\begin{gathered}
    0.85 \leq \cos \theta_{e} < 0.90 \;\cap\; 1.0 \leq E_{e} (\mathrm{GeV})< 1.65  \\
    0.90 \leq \cos \theta_{e} < 0.94 \;\cap\; 1.0 \leq E_{e} (\mathrm{GeV})< 2.0\;\label{eq:phasespace} \\
    0.94 \leq \cos \theta_{e} < 0.97 \;\cap\; 1.4 \leq E_{e} (\mathrm{GeV})< 3.0\; \\
    0.97 \leq \cos \theta_{e} \leq 1.00 \;\cap\; 1.4 \leq E_{e} (\mathrm{GeV})< 6.0\;
  \end{gathered}
\end{equation}

A $\chi^{2}$ minimization procedure~\cite{James1994MINUITFM}  fits the normalization of the signal and two background templates as described previously. A covariance matrix encodes the correlations between templates across all kinematic and template bins using systematic uncertainties which affect the shape of the ElectronID templates and the reconstructed electron kinematics within the fit. The fit produces a $\chi^{2}$ of 131 for 141 degrees of freedom. Using simulation with the best-fit parameters applied the sample purity is $13\pm1.3$\%. 
 Uncertainties on the purity prediction rise to 1.5\% in kinematic regions where the \numu+\numubar and NC normalization parameters become highly correlated due to templates no longer being as discriminating. The contribution of \nuebar CC background is removed from the renormalized signal template using the predicted contribution rate of 5\% from simulation. The resulting sample is estimated to contain  $9,200 \pm 1,000$ \nue CC interaction candidates, selected with 35\% efficiency. The signal sample is estimated to be 28\% QE, 20\% MEC, 31\% RES, 20\% DIS, and 1\% CC coherent pion production.

The flux-averaged double-differential \nue CC cross section in final state electron kinematic variables is constructed using
\begin{equation}
\left(\frac{d^{2} \sigma}{d \cos \theta d E}\right)_{i} = \frac{ \sum_{j} U_{ij}^{-1} N_{\mathrm{\nu_{e}}}(\cos \theta, E)_{j} }{N_{\mathrm{t}} \;  \phi \; \epsilon(\cos \theta, E)_{i}  \; \Delta \cos \theta_{i} \; \Delta E_{i}}, \label{eq:wideeq}
\end{equation}

\noindent where $N_\mathrm{\nu_{e}}$ the estimated \nue CC events from the template fit. $U_{ij}$, an unfolding matrix used to relate the reconstructed value in bin $j$ to the true value in bin $i$. The data are iteratively unfolded according to d'Agostini's method~\cite{ref:DAGOSTINI1995487} using two iterations as implemented in RooUnfold~\cite{ref:Adye:2011gm}. The optimal number was found by minimizing the average mean square error~\cite{james2006} calculated across simulated samples 
with random variations of systematic uncertainties. $N_t$ the number of nuclear targets in the fiducial volume, $\phi$ the integrated neutrino flux, $\epsilon$ the selection efficiency correction factor, and $\Delta E$ and $\Delta \cos \theta$ are the bin widths used for the electron kinematic variables. Bin widths are chosen small enough to match the detector resolution and large enough to include a statistically significant event sample. The average $E_{e}$ resolution is 350 MeV, and the angular resolution ranges from \unit{2}{\degree} for forward-going electrons to \unit{11}{\degree} for less forward-going electrons.

\Cref{table:uncertainties} summarizes the effects of various sources of uncertainty on the measurements. Systematic uncertainties are evaluated by varying the parameters used to model the neutrino flux, neutrino-nucleus interactions ($\nu$-A), detector response, and re-extracting the differential cross section. The difference between the cross section extracted using the nominal simulation and that extracted using the simulation with a varied parameter is taken as the uncertainty due to each parameter. The procedure accounts for changes in the compositions of backgrounds, selection efficiency, and event reconstruction due to the variations considered.

The dominant sources of systematic uncertainty are from the neutrino flux and $\nu$-A predictions. Uncertainties on the flux arise from hadron production uncertainties~(9\%)~\cite{Aliaga:2016oaz} and beam optics modeling~(4\%). A custom set of NOvA-specific uncertainties~\cite{ref:nova_xsec_tune} and uncertainties available from the GENIE generator are used to assess uncertainties on $\nu$-A interaction modeling. Model parameters affecting RES, DIS, and MEC predictions have the largest impact and become dominant at $E_{e} >$ \unit{3.0}{\giga \electronvolt}.

Non-leading sources of uncertainty come from detector calibration and modeling. These sources become dominant for $\cos \theta_{e} < 0.94$ and $E_{e} < \;$\unit{1.5}{\giga \electronvolt}.  Minor sources of uncertainty, which include detector mass, integrated beam exposure, beam intensity modeling, and the modeling of diffractive (DFR) $\pi^{0}$ production, are combined in the ``Other" category of \Cref{table:uncertainties}.  DFR modeling uncertainties are evaluated by reweighting the default $\nu$-H NC interactions producing a $\pi^{0}$ prediction from GENIE to an estimate based upon the calculation by Kopeliovich \textit{et al.}~\cite{PhysRevD.85.073003,npwebsite} as a function of $E_{\nu}$ and the Bj\"{o}rken scaling variables.  The average uncertainty on DFR modeling is 2.6\% across all bins.

 \Cref{table:uncertainties} also shows the weighted average bin-to-bin correlations for each source of systematic uncertainty. The large bin-to-bin correlations from the flux uncertainty indicate that it mainly impacts normalization. Interaction modeling also exhibits strong correlations across all bins, due to a combination of the template fitting procedure and the model parameters, such as the axial mass from the RES model and DIS model parameters that impact the selection efficiency prediction.
 
\begin{table}
\centering
\caption{Fractional uncertainties and correlations, broken down by source. Averages are taken across all bins reported in this measurement, weighted by the measured cross section.}
\begin{tabular}{r|c|c}
Source & Avg. Uncertainty (\%) & Avg. Correlation\\ \hline
Flux & 10.3  & 0.90\\
$\nu$-A Model & 9.8 & 0.64\\
Calibration & 5.9  & 0.05\\
Detector Model & 5.6  & 0.21\\
Other & 2.8  & 0.03\\
Statistical & 7.4 & 0.02\\ \hline
Total & 18.2 & 0.59\\
\label{table:uncertainties}
\end{tabular}
\end{table}

\begin{figure}
\centering
\includegraphics[scale= 0.433]{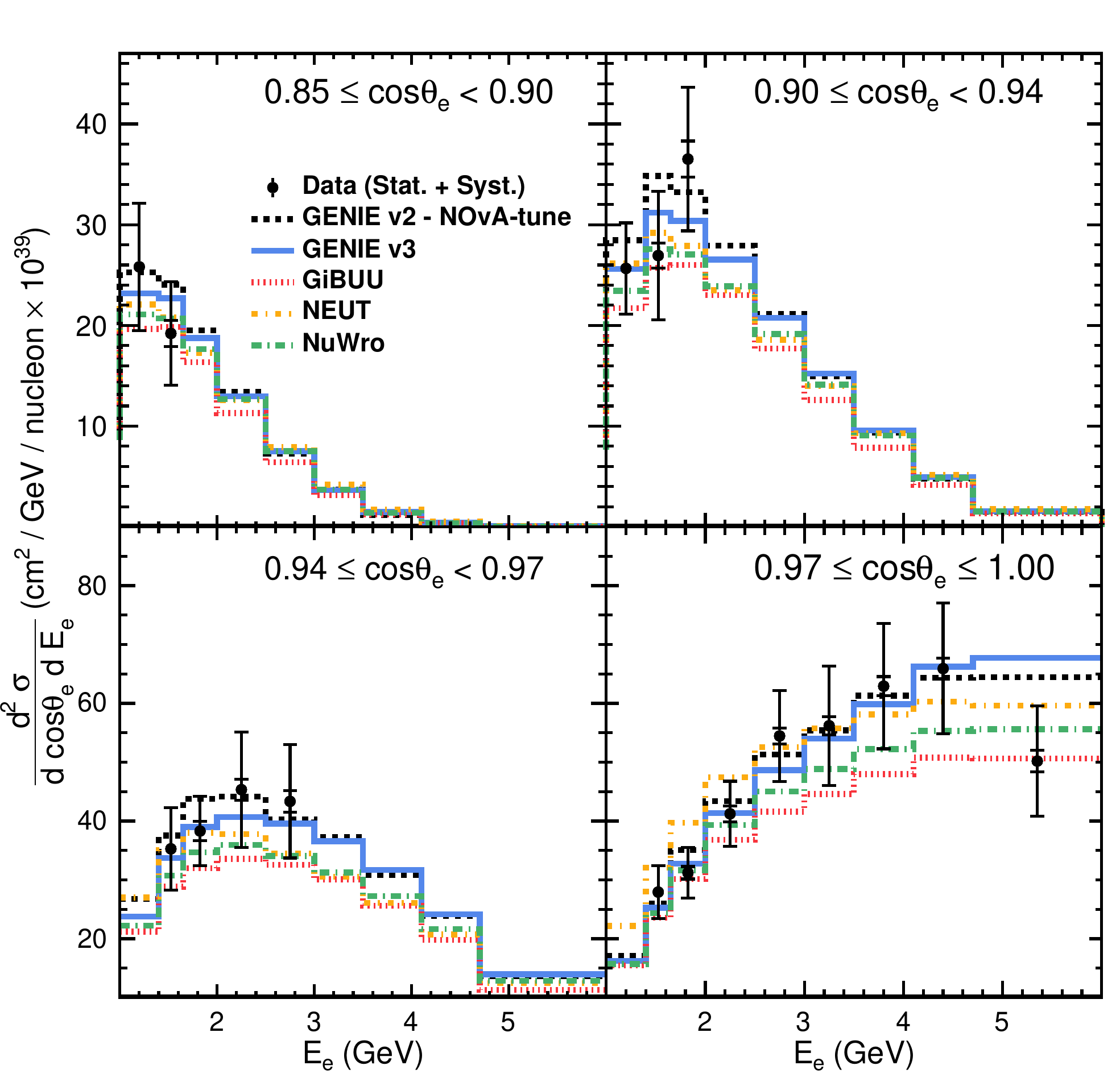}
\caption{Extracted double differential cross section, subdivided in slices of electron angle. The outer error bars of the data represent total uncertainties, while the inner error bars are statistical only. The data are compared to several models.}
\label{fig:2d_result}
\end{figure}

\begin{figure}
\includegraphics[scale= 0.433]{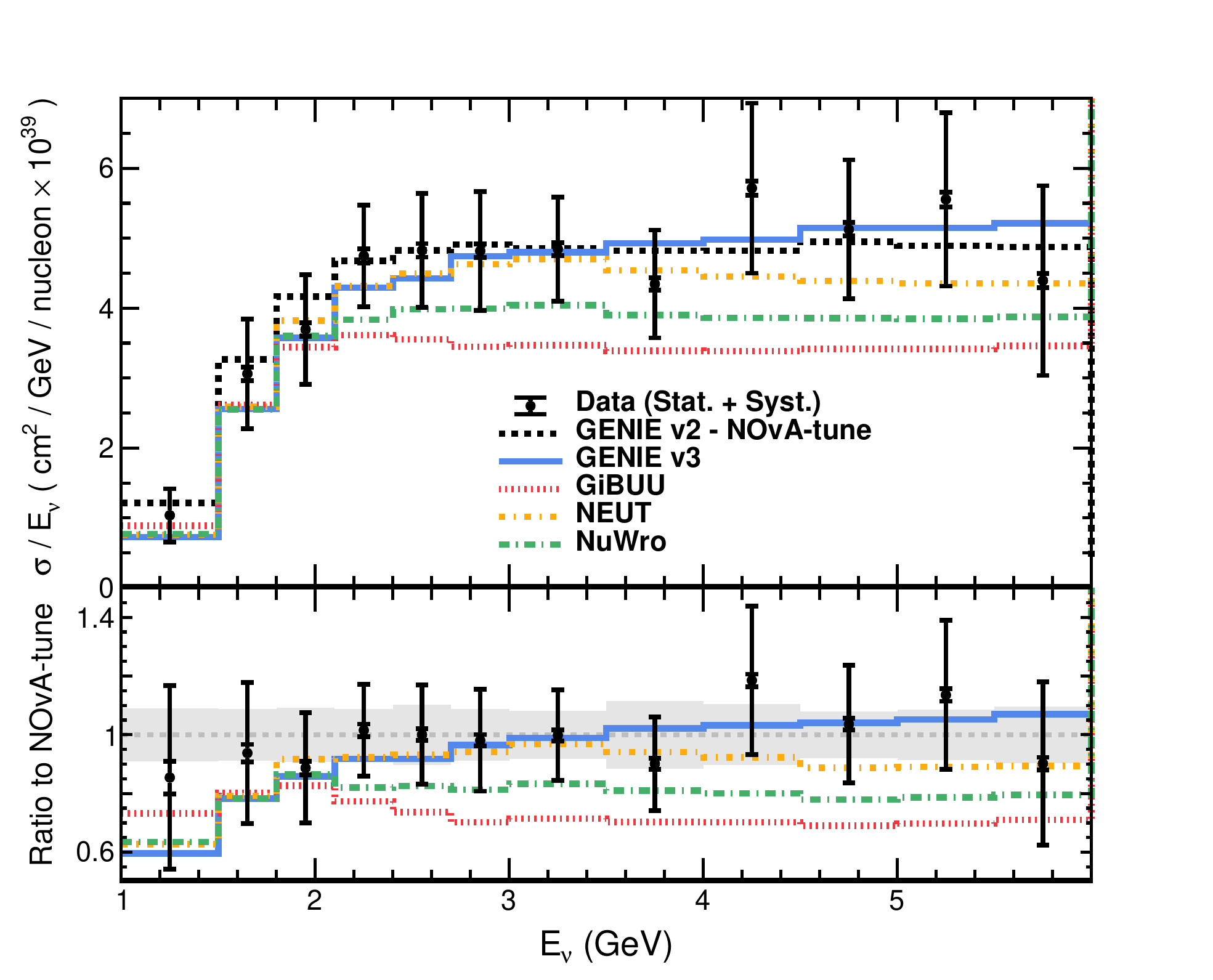}   
\caption{Top: The differential cross section vs.~$E_{\nu}$. Bottom: Comparisons as a ratio to the NOvA-tune prediction. The grey band represents the normalization uncertainty from the neutrino flux prediction. The data are presented showing statistical and total uncertainties.}
\label{fig:1d_enu}
\end{figure}

\begin{figure}
\includegraphics[scale= 0.433]{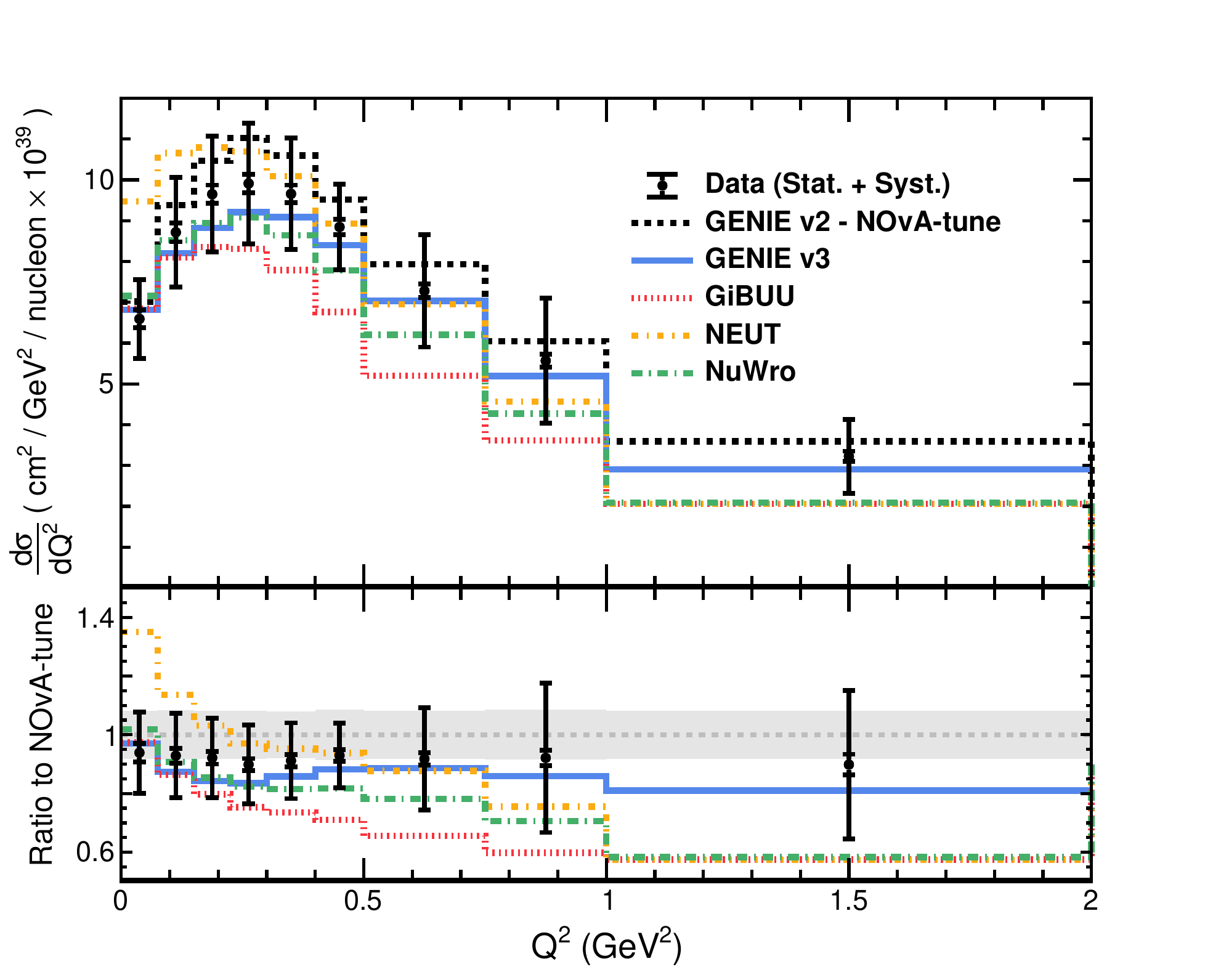}   
\caption{Top: The differential cross section vs. $Q^{2}$. Bottom: As in Fig. 4, but for $Q^{2}$. }
\label{fig:1d_q2}
\end{figure}

Three results are presented:~the flux-integrated double-differential cross section vs.~electron energy and angle shown in \Cref{fig:2d_result}, the cross section vs.~$E_{\nu}$ shown in \Cref{fig:1d_enu}, and the differential cross section vs.~$Q^{2}$ shown in \Cref{fig:1d_q2}. All results are calculated from data using unfolding and efficiency corrections derived from the NOvA-tune.  For the one-dimensional results we first apply the phase space restrictions in $E_{e}$ and $\cos \theta_{e}$ given in Eqn. 1. The result is then extracted with unfolding and efficiency corrections in $E_{\nu}$ or $Q^{2}$. 

The extracted cross sections are compared to predictions from several generators: GENIE models (NOvA-tune, v2.12.2, and v3.00.06 with a global configuration N18\_10j\_02\_11a), NuWro  2019~\cite{Golan:2012rfa}, GiBUU 2019~\cite{Lalakulich:2011eh}, and NEUT 5.4.0~\cite{Hayato:2009zz}. \Cref{table:chi2} summarizes $\chi^{2}$ values for each model compared to data with a full treatment of bin-to-bin correlations. The values show no model is consistently favored across the ensemble of measurements.

Though the GiBUU and NuWro predictions have normalizations systematically lower than the data in \Cref{fig:2d_result,fig:1d_enu,fig:1d_q2} (15\% and 10\%, respectively) their $\chi^{2}$s in \Cref{table:chi2} are comparable to other models' due to the phenomenon of Peelle's Pertinent Puzzle (PPP)~\cite{smith1991probability,refId0}. PPP arises when the dominant uncertainty of a result is from highly correlated normalization uncertainties, like the flux uncertainty in \Cref{table:uncertainties}. Under these circumstances, the best-fitting model as reckoned by $\chi^{2}$ can be well outside the data points. On the other hand, disagreements with the GENIE predictions are dominated by the discrepancy in the  $E_{e} > $ \unit{4.75}{\giga \electronvolt} bin and its correlations with the other bins. When this bin is excluded from the calculation, the $\chi^{2}$'s for the GENIE predictions approach those of the NuWro and GiBUU predictions. NEUT shows slight disagreement throughout that angular slice, predicting a softer energy spectrum than is observed in data.

In $E_{\nu}$ comparisons NuWro and GiBUU plateau at a lower total cross section than the data as described above. The differential-cross section vs.~$Q^{2}$ shows preference towards the NOvA-tune and GENIE v3 predictions. NEUT predicts a softer $Q^{2}$ distribution than seen in data, while GiBUU and NuWro accurately predict the low $Q^{2}$ behavior but tension is seen at high $Q^{2}$. The grey band in \Cref{fig:1d_q2} shows the size of the flux-related normalization uncertainties, which illustrates that differences seen between data and the generators are not consistent with an overall change in the cross-section normalization.

\begin{table}
\centering
\caption{$\chi^{2}$ calculated for each model presented in \Cref{fig:2d_result,fig:1d_enu,fig:1d_q2} compared to data. Degrees of freedom are 17, 12, and 9 for the double-differential, $E_{\nu}$, and $Q^{2}$ results, respectively.}\label{table:chi2}
\begin{tabular}{r|c|c|c}
Generator      & $\frac{d^{2}\sigma}{d \cos \theta_{e} d E_{e}}$& $\sigma(E_{\nu})$ & $\frac{d\sigma}{d Q^{2}}$ \\ \hline
GENIE v2 - NOvA-tune & 24.1 & 13.4 & 1.3  \\                  
GENIE v2.12.2        & 24.3 & 14.3 & 19.6 \\ 
GENIE v3.00.06       & 27.4 & 21.6 & 3.4 \\                  
GiBUU 2019           & 17.5 & 16.0 & 14.7 \\
NEUT 5.4.0           & 25.1 & 16.9 & 45.0 \\
NuWro 2019           & 18.7 & 15.3 & 10.0 \\                                 
\end{tabular}
\end{table}

This paper presents the first measurement of the inclusive charged-current double-differential electron-neutrino cross section vs. electron energy and angle, using the NOvA near detector. This provides new information concerning directly observable final-state electron kinematic variables necessary for neutrino energy estimation and efficiency correction in \nue appearance measurements. Measured cross sections are shown compared to multiple versions of GENIE, and to GiBUU, NEUT, and NuWro event generators, in the neutrino energy range from 1 to \unit{6}{\giga \electronvolt}. The models show general agreement with the data in the two-dimensional phase space as well as various levels of success in the neutrino energy and $Q^{2}$ measurements. Disagreements are seen in both overall normalization across all reported measurements and with certain shape differences, primarily seen in $Q^{2}$. The data related to this measurement and systematic covariance matrices can be found at \cite{ref:data_release}.

\section{Acknowledgements}
\begin{acknowledgments}
This document was prepared by the NOvA collaboration using the resources of the Fermi National Accelerator Laboratory (Fermilab), a U.S. Department of Energy, Office of Science, HEP User Facility. Fermilab is managed by Fermi Research Alliance, LLC (FRA), acting under Contract No. DE-AC02-07CH11359. This work was supported by the U.S. Department of Energy; the U.S. National Science Foundation; the Department of Science and Technology, India; the European Research Council; the MSMT CR, GA UK, Czech Republic; the RAS, MSHE, and RFBR, Russia; CNPq and FAPEG, Brazil; UKRI, STFC and the Royal Society, United Kingdom; and the state and University of Minnesota.  We are grateful for the contributions of the staffs of the University of Minnesota at the Ash River Laboratory, and of Fermilab. For the purpose of open access, the author has applied a Creative Commons Attribution (CC BY) license to any Author Accepted Manuscript version arising.
\end{acknowledgments}

\bibliographystyle{apsrev4-1}
\bibliography{nova_nuecc_inc_dd.bib}

\end{document}